# Metasurface-Enabled On-Chip Multiplexed Diffractive Neural Networks in the Visible

Xuhao Luo[1,#], Yueqiang Hu[1,4,#,*], Xin Li[1], Xiangnian Ou[1], Jiajie Lai[1], Na Liu[2,3], and Huigao Duan[1,4,*]

[1] National Research Center for High-Efficiency Grinding, College of Mechanical and Vehicle Engineering, Hunan University, Changsha 410082, P.R. China

[2] 2nd Physics Institute, University of Stuttgart, Pfaffenwaldring 57, 70569 Stuttgart, Germany.

[3] Max Planck Institute for Solid State Research, Heisenbergstrasse 1, 70569 Stuttgart, Germany.

[4] Advanced Manufacturing Laboratory of Micro-Nano Optical Devices, Shenzhen Research Institute, Hunan University, Shenzhen, 518000, China

[#] These authors contributed equally.

[*] Corresponding authors. Email: huyq@hnu.edu.cn, duanhg@hnu.edu.cn

## Abstract

Replacing electrons with photons is a compelling route towards light-speed, highly parallel, and low-power artificial intelligence computing. Recently, all-optical diffractive neural deep neural networks have been demonstrated. However, the existing architectures often comprise bulky components and, most critically, they cannot mimic the human brain for multitasking. Here, we demonstrate a multi-skilled diffractive neural network based on a metasurface device, which can perform on-chip multi-channel sensing and multitasking at the speed of light in the visible. The metasurface is integrated with a complementary metal oxide semiconductor imaging sensor. Polarization multiplexing scheme of the subwavelength nanostructures are applied to construct a multi-channel classifier framework for simultaneous recognition of digital and fashionable items. The areal density of the artificial neurons can reach up to $6.25\times10^{6}/\mathrm{mm}^{2}$ multiplied by the number of channels. Our platform provides an integrated solution with all-optical on-chip sensing and computing for applications in machine vision, autonomous driving, and precision medicine.

## Introduction

Artificial intelligence (AI) is a technology for simulating and extending human intelligence[1,2], of which artificial neural networks (ANN) is one of the most widely used frameworks implemented in electronic equipment to digitally learn the representation and abstraction of data for performing advanced tasks[3,4]. ANN enables rapid performance improvement of single specific tasks, such as image recognition[5,6], speech recognition[7], natural language processing[8], among others[9-13]. However, the human brain works as a multi-channel system [14] including sight, hearing, smell, taste, and touch as shown in Figure 1a, and even each channel contains multiple sub-channels. Therefore, to achieve human-like artificial general intelligence (AGI), different capabilities should be multiplexed in a single AI system for multi-skilled AI that has wide application potential in smart home, autonomous driving, and somatosensory interaction. Meanwhile, multiplexed AI systems can greatly increase the computing scale and degree of parallelism.

Recently, optical neural networks[15-21] have attracted much attention due to their high speed, high parallelism and low energy consumption compared with neural networks running by electrons. As a kind of optical neural network, the all-optical diffractive neural network has been proposed by constructing physical multiple diffractive layers to simulate neural layers to achieve specific functions[22-27]. It inherits the advantages of the optical neural networks and has the property of low-cost scalability. Nevertheless, the existing diffractive neural network devices, like conventional neural networks, cannot perform multiplexed information processing[28-32]. In addition, they are usually implemented in large wavelength bands with bulky sources and detectors, which cannot be directly integrated with low-cost imaging sensor chips for miniaturization.

Here, we demonstrate a multiplexed metasurface-based diffractive neural network (MDNN) integrated with a complementary metal-oxide semiconductor (CMOS) imaging sensor for on-chip multi−channel sensing in the visible range. Metasurfaces are novel planar optical elements consisting of subwavelength resonators for manipulating the wavefront of light[33]. The unprecedented ability of metasurfaces for multiparametric modulation makes them a powerful platform for multifunctional multiplexing in a single element[34-36]. We demonstrate multitasking at the speed of light through polarization multiplexed metasurface, using a plane wave of the amplitude or phase of the object to be recognized as the input signal to achieve simultaneous recognition of digital and fashionable items. The multi-channel classifier framework is constructed by computer machine learning based on an error backpropagation approach. Due to ultra-flat and ultra-thin characteristics of metasurfaces, integration of the MDNN with CMOS chip is achieved which provides the possibility of high-volume manufacturing in semiconductor plants with the CMOS-compatible processes. This is the first on-chip all-optical diffractive neural network realized in the visible range using metasurfaces. The areal density of neurons is greatly enhanced due to the subwavelength structure and proportional to the number of channels.

## Results

The framework of MDNN for multiplexed classification shown in Figure 1b, comprises different types of targets to be recognized in multiple channels as inputs (e.g., handwritten digits, fashion items, letters and so on), hidden layers with meta-neurons encoding multiplexed phases, and detectors with sub-areas for multi-channel detection. A training principle similar to that of conventional electronic neural networks is employed for each channel, which generally consists of three components: a single input layer, a hidden layer with at least one layer of neurons, and a single output layer.

By deep learning with error back propagation, the multidimensional phase distributions are iteratively updated such that eventually the information from different channels converges to their specific detection regions, each corresponding to an identification class. The object can be encoded as an amplitude or phase component, propagated and modulated at the speed of light in meta-neurons. To achieve phase encoding of multiple channels for meta-neurons, as a proof-of-concept, we demonstrate here a kind of architecture based on polarization-multiplexed metasurfaces[37,38] (see Figure 1c). Each hidden layer consists of asymmetric meta-units, enabling the birefringence properties. By tuning the structural parameters of each meta-unit, polarization-dependent phase responses can be encoded. This allows parallel multitasking through different polarization incidence of targets. Moreover, due to the planar nature of the metasurface, it is easy to integrate it into a CMOS imaging sensor to realize an on-chip integrated AI chip.

The basic physics of the hidden layer design consisting of polarization-multiplexed meta-units is discussed. According to the Huygens-Fresnel principle [39], each point on the wave-front can be regarded as the source of the secondary spherical wave, and the shape of the new wave-front at the next moment is determined by the envelope of the secondary spherical wave. As such, each meta-unit in a particular polarization state can be considered as a neuron (i.e., a monopole source) fully connected to the preceding and following neurons. Based on the Rayleigh-Sommerfeld diffraction integral[40] and Jones matrix Fourier optics[41], the $l+1$ layer optical field of the all-optical meta-neurons network can be expressed as

$$\boldsymbol{U}(\vec{r}^{l+1}) = \iint_{-\infty}^{+\infty} \boldsymbol{U}(\vec{r}^{l}) * \tilde{J}_{meta}(\vec{r}^{l}) * h(\vec{r}^{l+1} - \vec{r}^{l})dxdy \qquad (1)$$

where $\boldsymbol{U}(\vec{r}^{l})$ is the light field irradiated to the $l$th layer, and for $l=1$ , $\boldsymbol{U}(\vec{r}^{l})$ is the projected light of the object to be identified. And $\tilde{J}_{meta}(\vec{r}^{l})$ is the Jones matrix of the birefringent metasurface of the $l$th layer, which can be expressed by $\tilde{J}_{meta}(\vec{r}^l) = \Gamma(\theta(x,y))\begin{bmatrix} a_x(x,y)e^{j\varphi_x(x,y)} & 0 \\ 0 & a_y(x,y)e^{j\varphi_y(x,y)} \end{bmatrix}\Gamma(-\theta(x,y))$, whereas $\theta = 0$. And $h(\vec{r}^{l+1} - \vec{r}^l) = \frac{1}{2\pi}\frac{z^{l+1}-z^l}{R}\left(\frac{1}{R} - jk\right)\frac{e^{jkR}}{R}$ is the first Rayleigh-Sommerfeld impulse response function, where $R = \sqrt{\left(x_p^{l+1} - x_i^l\right)^2 + \left(y_p^{l+1} - y_i^l\right)^2 + (z^{l+1} - z^l)^2}$. Thus, the forward propagation model of MDNN is constructed, by a cross-entropy loss function and a stochastic gradient descent approach to achieve desired output via the network. Detailed model training and derivation are demonstrated in "Methods" section and Supplementary Note 1.

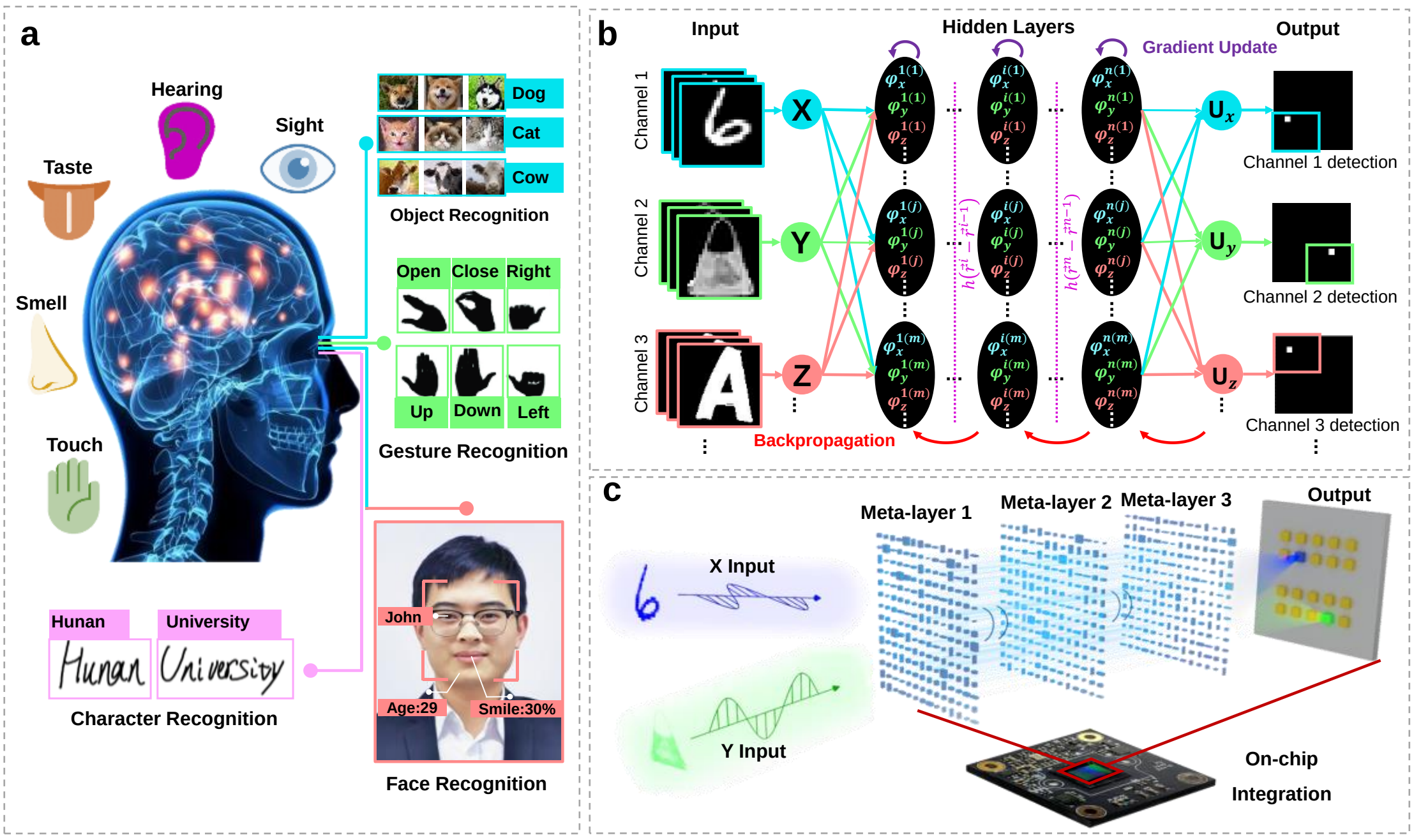

***Figure 1. Schematic of multiplexed metasurface-enabled diffractive neural networks (MDNN) integrated on an imaging sensor chip.*** *(a) Multi-channel senses of human brain mainly comprise sight, hearing, smell, taste and touch, among which vision can be subdivided into object recognition, gesture recognition, character recognition and face recognition, etc. (b) Architecture of the MDNN. The meta-neurons of the multiple networks are trained individually to obtain multiplexed phase distributions, which are*

*optimized by an error backpropagation algorithm using computer. (c) Optical layout of polarization-dependent object classification for the MDNN concept. The input is the light carrying information about the object to be recognized, e.g., a handwritten digit or a fashion product in x- and y- polarization, respectively. The hidden layers consist of polarization-multiplexed metasurfaces acting as neurons, and subsequently converge the diffraction energy to the corresponding photoelectric detection region on the CMOS chip (i.e., the output layer of network).*

Figure 2 schematically shows the detailed design of the polarized-multiplexed dual-channel meta-units for the MDNN. The metasurface is composed of subwavelength rectangular $TiO_2$ nanopillars. A nanopillar with two independently tunable structural parameters $(D_x, D_y)$, a fixed height $H$ and a period $p$, is delineated in Figure 2a. Its rectangular cross section leads to different effective refractive indices along the two crossed axes, which is the fundamental mechanism for achieving polarization multiplexing. When linearly polarized light is incident along the corresponding axes, the nanopillar produces polarization-dependent phase shifts which can be expressed as a function of $D_x$and $D_y$. The amplitude and phase under $x$- and $y$-polarization are simulated by the finite-difference time-domain (FDTD) method, where the wavelength is chosen to be 532 nm and $p$ is set to 400 nm (Figure 2b-e). The nanopillars have a height $H$ of 600 nm without cladding to achieve a combination of multiplexed phases covering approximately two 0-2π ranges as well as a high transmittance (More details about the nanopillars with cladding are in Supplementary Note 2 and Figure S1.). Figures 2f-g demonstrate top-view and oblique scanning electron microscopy (SEM) images of the fabricated $TiO_2$ metasurface (see more details about fabrication of the metasurface in "Methods" and Supplementary Note 9). We designed and fabricated a dual-focus metalens to verify the polarization-multiplexing, and the simulation and experimental results agree well, as shown in Figures 2h-i.

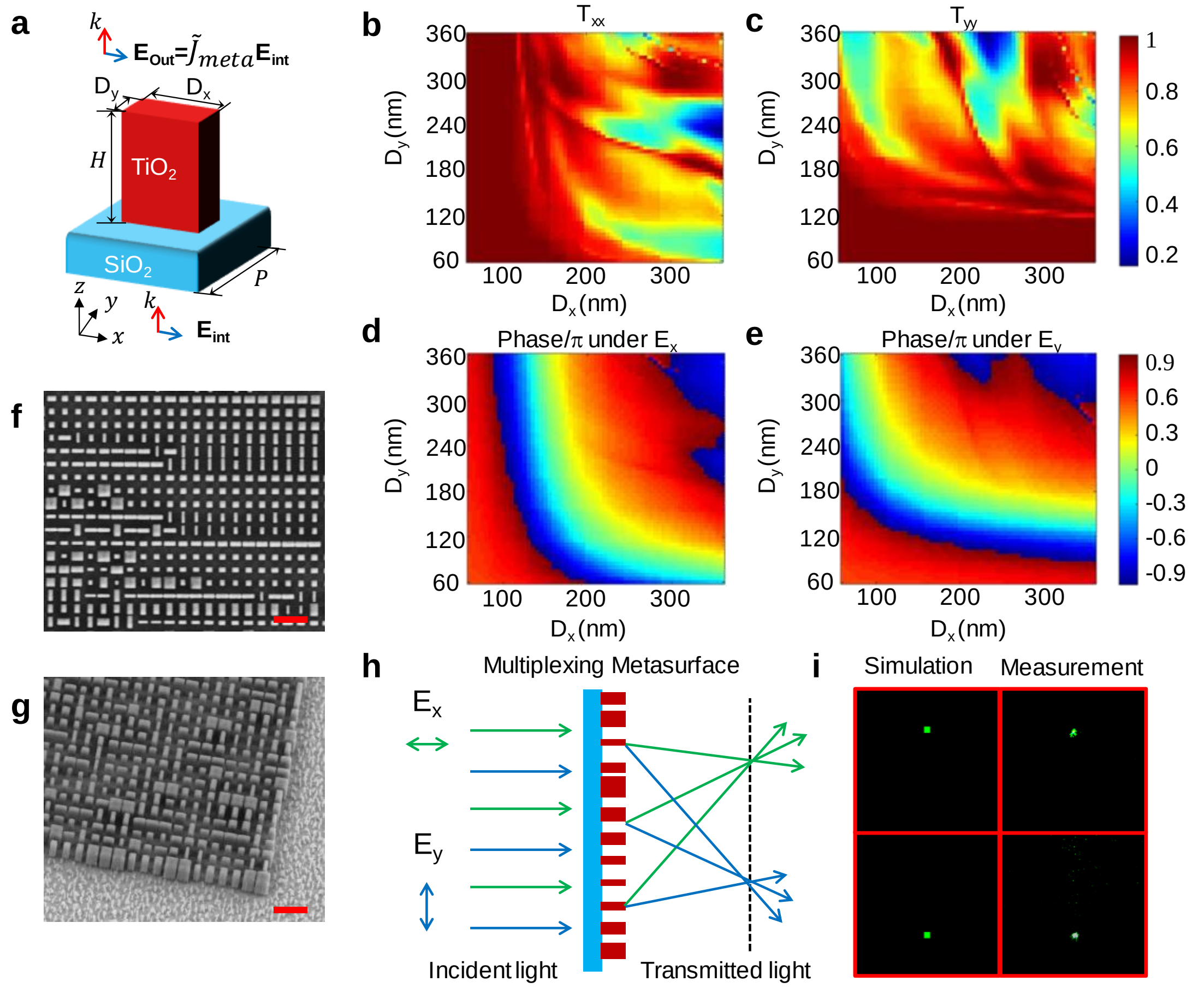

***Figure 2. Design of the multi-channel meta-neurons.*** *(a) Schematic of a single $TiO_2$ meta-unit with a fixed height H, while tunable structure dimensions $D_x$ and $D_y$. Each meta-unit acts as a neuron that has multiplexed phase profiles trained by machine learning. (b-e) Simulated values of the transmission coefficients ($T_{xx}$, $T_{yy}$) and the phase shifts ($\varphi_{xx}$,$\varphi_{yy}$) under x- and y-polarized optical waves, respectively. An incident wavelength of 532 nm, a nanopillar period of 400 nm and a height of 600 nm are assumed. SEM images of the fabricated metasurface: (f) top and (g) side views. The scale bar is 1 μm. (h) Bifocal metasurfaces for demonstrating independent phase modulations of the designed meta-neurons under the two orthogonal polarizations. (i) Simulated and experimental results of the focal intensity field. The green arrow refers to x polarization, and the blue one refers to y polarization.*

To demonstrate a polarization-multiplexing MDNN for multi-channel identification, a dual-channel neutral network was trained from the MNIST[42] (Modified National Institute of Standards and Technology) and the Fashion-MNIST[43] datasets. The training convergence with respect to the epoch number is shown in Figure 3a. The physical plane of the network's output is divided into 10 discrete detection regions, each representing one class of the datasets, and the region presenting the highest intensity implies the class of the object being recognized. It is obvious that the recognition accuracy can be improved with the increase of the number of hidden layers. The accuracy is slightly higher for MNIST than that of Fashion-MNIST (inset in Figure 3a), probably because the data complexity of the former is lower than that of the latter. Figure 3b summarizes the effect of the number of neurons in a hidden layer, and the relative size of individual detection region on the accuracy, where the network utilizes three hidden layers with 28×28 neurons per layer and a period of 400 nm, taking the MNIST data as an example. It is found that the small detection region helps to slightly improve the recognition accuracy. This will help to enable detection of more sub-channels in a fixed sensor area. However, the number of neurons does not have much effect on the accuracy, mainly because when preprocessing the input data, we perform an isometric scaling of the input, which does not affect the amount of input information, but only increases the number of neurons in the hidden layers (more details in Figure S2). Since MDNN differs from conventional diffraction networks in that the meta-units introduces additional amplitude modulation as shown in Figure 2b, c, the phase-only and amplitude-crosstalk networks are compared to analyzed the effect of the amplitude (Figure 3c). Taking the handwritten digit "3" as an example (see more examples in Figure S3), for both networks, the three-layer all-optical neural network we designed can accurately redistribute the input energy to the detection region as expected. When we take the amplitude-crosstalk of the metasurface into account in the computation of the phase-only network with 10,000 handwritten digits testing dataset (The comparison of these two networks based on Fashion-MNIST is presented in Figure S5), the obtained recognition results have negligible error effects (Figure 3d-f). As demonstrated in Figure 3g, the normalized distribution of the energy in the respective target detection regions is obtained by collating all test data for handwritten digits from “0” to “9” with amplitude-crosstalk, where the error bars show the difference compared to the phase-only network. It can be seen that the average energy distribution of each target reaches more than 25%. The effect of the amplitude-crosstalk on the energy distribution is negligible, and the underlying reason is that the phase plays a major role in the modulation of light by the metasurface.

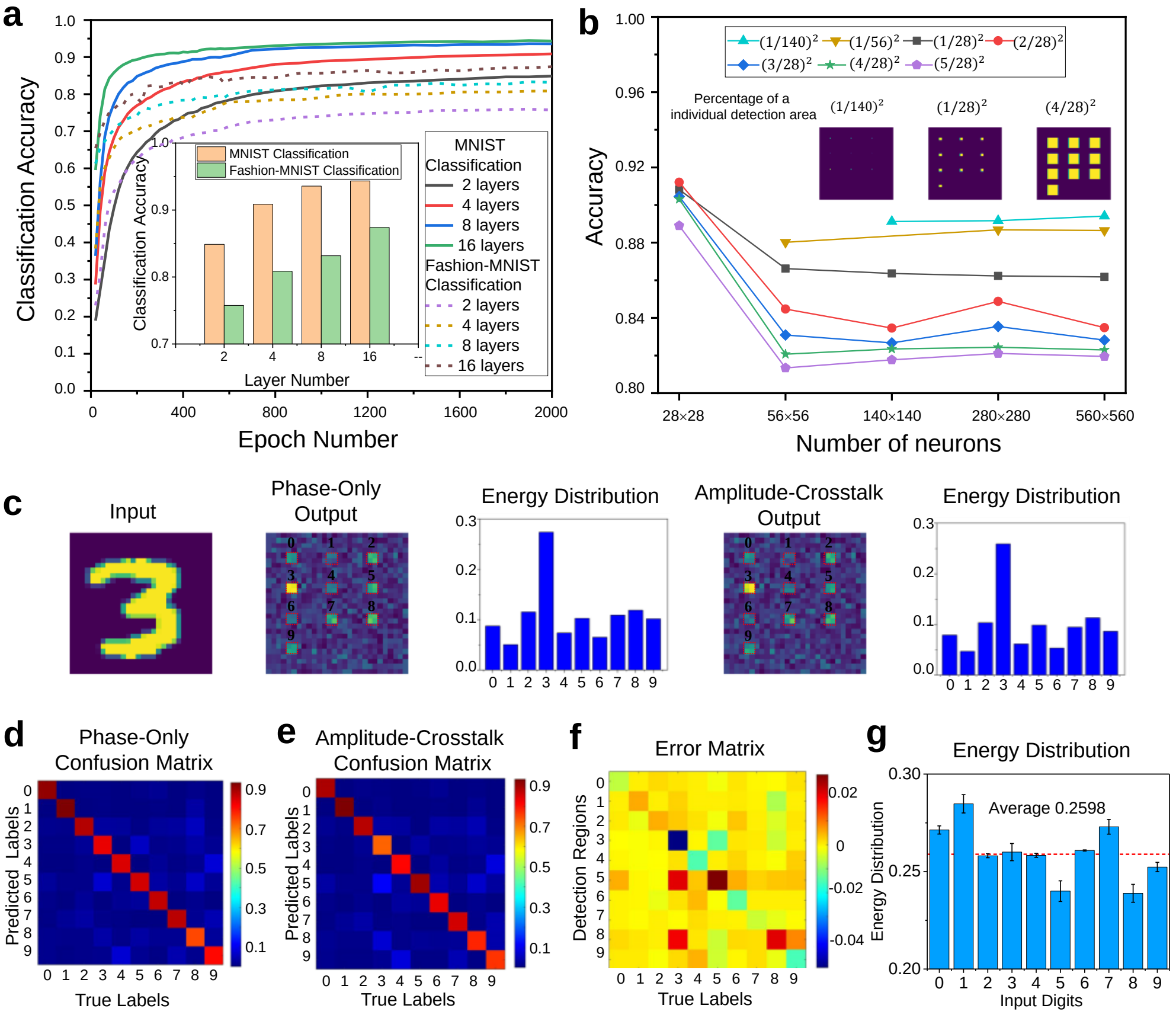

*__Figure 3. Numerical demonstration of the MDNN.__ (a) Training convergence plots for phase-modulated MNIST and Fashion-MNIST on the epoch number. Insets:*

*performance comparisons of these two classifiers, reporting that both have relatively close classification accuracy and that the accuracy can be further improved as the number of hidden layers increases. (b) Classification accuracy as a function of the number of neurons for MNIST dataset, with different sizes of detection regions. (c) Simulated output patterns and energy distribution percentages in the detection plane corresponding to a handwritten input of "3", in the case of phase-only and amplitude-crosstalk networks, respectively. The confusion matrixes for (d) phase-only (amplitude is considered to be a constant value) and (e) amplitude-crosstalk MDNN are demonstrated, where the amplitude-crosstalk is based on the complex amplitude of the meta-unit in Figure 2. (f) Percentage error matrix between (d) and (e). (g) Percentage of energy distribution for each digit of the network with amplitude-crosstalk. The error bars show the differences in comparison with the phase-only network.*

Since the metasurfaces are subwavelength arrayed devices, scalar diffraction theory is no longer applicable in principle, due to its disregard of polarization properties and inter-structural interactions. To further verify the functions of MDNN, we also performed 3D full vector simulation by FDTD methods. The processes of scalar and vector simulation are compared in Figure 4a. The scalar simulation is to calculate the light wave as a scalar quantity, which is an approximation of the actual propagation process, while the vector simulation can perfectly reproduce the interaction process between the light wave and the metasurface to obtain the information of the propagation, intensity and power. First, the input object to be detected is the amplitude or phase distribution of the polarization source; then the multi-channel diffraction phase is calculated by deep learning. For scalar simulation, the phase is directly substituted into the diffraction integral for a layer-by-layer calculation to obtain the output. For vector simulation, the phase distribution is transformed into the structural parameters of the corresponding $i$-layer metasurface array, followed by FDTD simulation to obtain the near field, and then the far field is extrapolated. If the last layer of meta-neurons calculation is completed, the output light intensity distribution is obtained. The dual-channel all-optical neural network is trained based on two hidden layers, and the MNIST and Fashion-MNIST training convergence is shown in Figure S2, indicating that both networks achieve a high accuracy rate of $> 99\%$. Figure 4b shows the scalar diffraction calculation and vector simulation for a set of polarized dual-channel dual-object recognition (More examples are presented in Supplementary Note 7). The light propagation in the $z$-direction from the last meta-layer to the output plane for the handwritten digit "3" is illustrated in Figure 4c. It can be seen that the MDNN can accurately focus the input energy on the target detection region for each channel in vector simulation. Figure 4d shows the focused light field curves of the four identified objects obtained in the $x$-axis of the intercepted detection region. The peak intensities of all field intensities appear in the regions corresponding to the classified targets, in agreement with expectations. Figure 4e gives the recognition normalized energy distribution of the 10 sets of vector simulations obtained from the same simulation step, and it is obvious that the average percentage of energy for the classified targets are all higher than 80%, indicating that this FDTD vector simulation also verifies the effectiveness of MDNN.

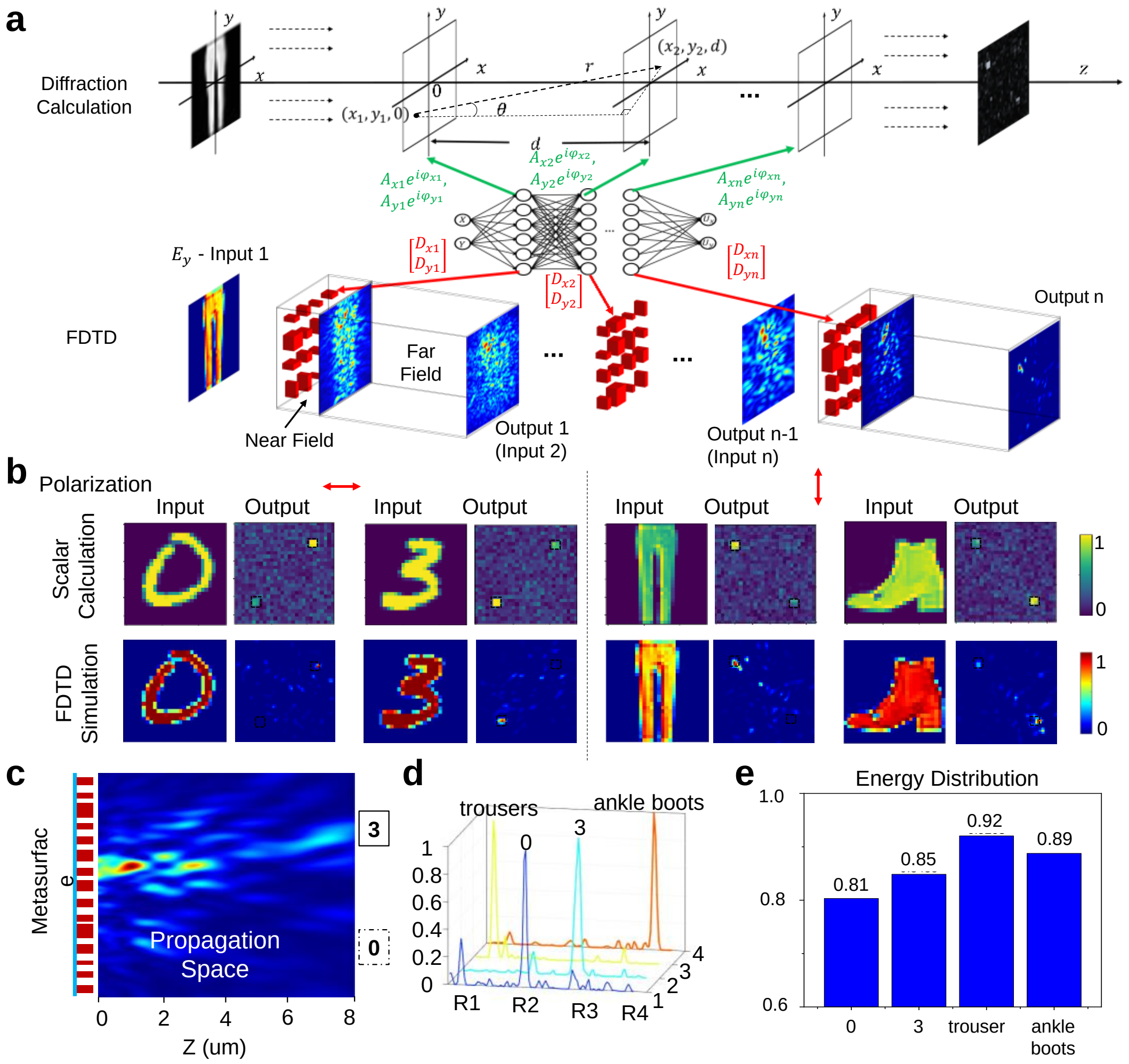

***Figure 4. Vector simulations of MDNN.*** *(a) Flow charts of scalar diffraction calculation and vector FDTD simulation. (b) Comparison of simulation results between scalar diffraction calculation and vector FDTD calculation for multi-channel classification. (c) The electric field distribution in the z-plane simulated by the handwritten input of "3" in (b), demonstrates that the light propagation is focused into the target region. (d) The output intensity in (b) is normalized along the x-direction distribution, and the maximum peaks are all confined to the detection region. (e) The average energy distribution of simulated 10 groups for the four types of objects, all of which are randomly selected in the MNIST and Fashion-MNIST datasets, reached more than 80%.*

According to the above design, we integrate the fabricated metasurfaces with CMOS imaging sensor, demonstrating dual-class object recognition within dual channels, as a proof-of-concept. Figures 5a-b show a schematic and physical diagram of an on-chip MDNN to achieve classification tasks in visible light. To simplify the experiment, the processed binary Aluminum mask is utilized as the input amplitude of the MDNN, i.e., where the position without (with) Aluminum structure can (cannot) transmit light with amplitude of 1 (0). The Aluminum mask, spacer, and polarization-multiplexed metasurface are integrated on the substrate by an electron beam lithography (EBL) overlay process, and then are monolithically adhered onto the COMS imaging sensor by an optically clear adhesive (OCA) with a thickness of 100 μm for On-chip integration. Figure 5c illustrates a zoomed-in view of a single integrated metasurface under a microscope with a handwritten digit "0" mask as an example. Figure 5d shows the two sets of Aluminum masks we fabricated (e.g., handwritten digits “0” and “1”, and the fashion products t-shirts and sneakers), and the device after adding the pacer, where the spacer is to protect the Aluminum mask. The polarization-multiplexed dual-channel neural networks were trained for a single hidden layer with 280 × 280 meta-neurons (78400 in total, and an area size of 112×112 $\mu m^2$) in the experiments, and the training convergence of MNIST and Fashion-MNIST with respect to epoch number is shown in Figure S4, where both networks achieved a higher accuracy of greater than or equal to 99%. The phase distribution obtained after training under dual polarization is presented in Supplementary Note 5. To test the dual-channel performance and to measure the accuracy of the experimental classification, we compare the output of simulations and experiments for two-class objects in x- and y- polarization, respectively. Figure 5e exhibits the simulation results and experimental results of an all-optical neural network based on an on-chip polarization-multiplexed metasurface. The characterization setup to perform the functions of MDNN can be found in Figure

S9. Before the integration with CMOS chip, by varying the diffraction distance (0 to 100 μm), the output images of the MDNN are detected by the charge-coupled device (CCD) camera (see Supplementary Movie S1-4). The final device demonstrates the ability to identify dual targets in a dual channel, where the maximum energy is accurately clustered in the target detection region. Note that the output images of device characterization are collected directly by the CMOS imaging sensor. The experimental classification results match 96% with the simulation. Compared with the simulated results, the intensity of the experimental target detection region is slightly lower while the rest of the region is relatively higher, which is mainly due to the error of the polarizer that cannot completely eliminate the orthogonally polarized light.

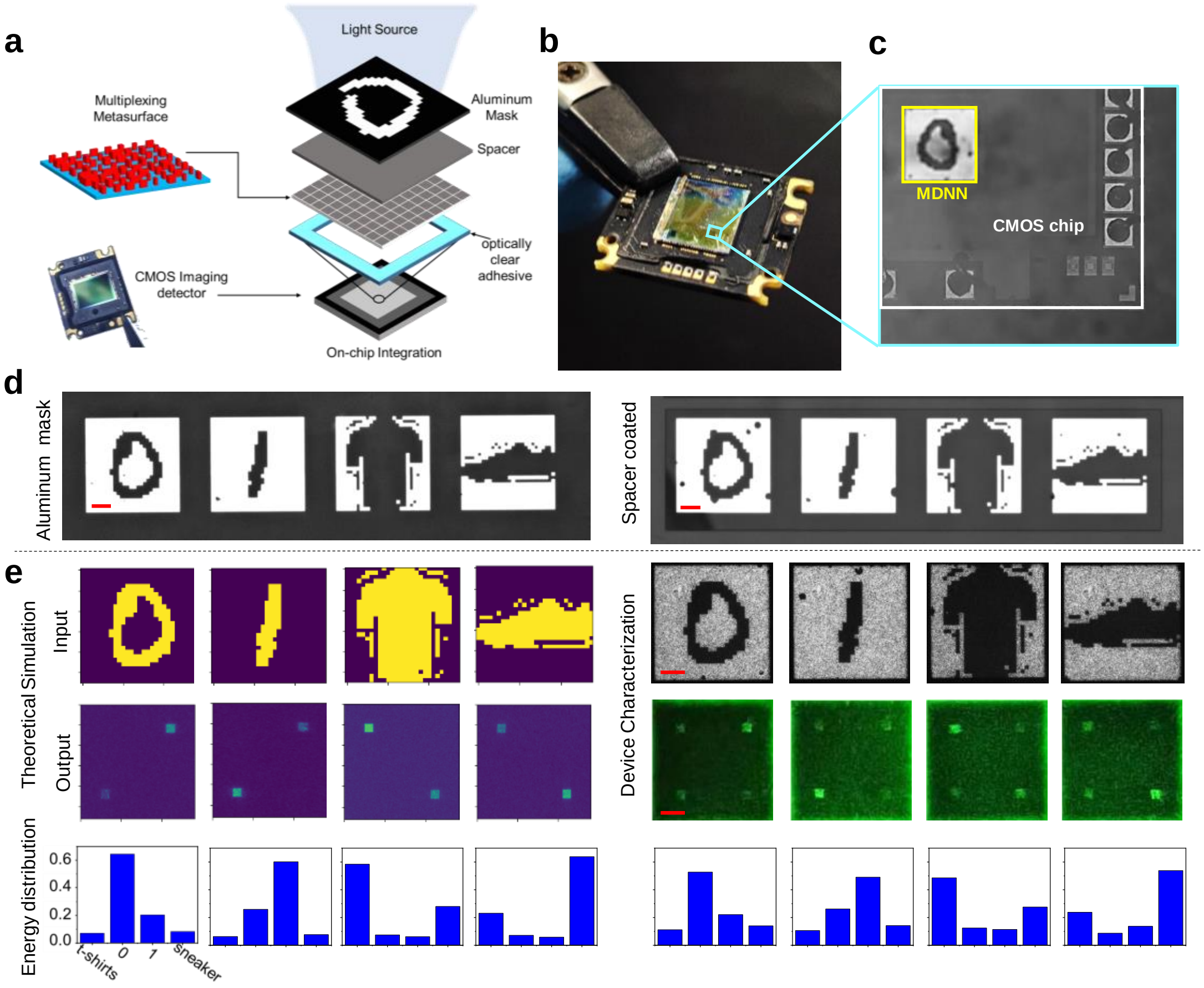

***Figure5. Experimental demonstration of the on-chip MDNN.*** *(**a**) Exploded schematic diagram of the MDNN integrated with a CMOS chip. (**b**) Physical photograph of the* *on-chip MDNN. (**c**) Optical micrograph of a single fabricated MDNN built on a CMOS imaging sensor. (**d**) Mask formed after aluminum deposition (left) and after adding a spacer layer (right). (**e**) left: Simulation inputs and outputs of the designed MDNN, and their energy distribution; right: the final metasurfaces-based device (top), The output field intensity detected by an imaging sensor for these cases (middle), and energy distribution of the experiment results (bottom). The scale bar is 20 $\mu m$.*

**Conclusion**

We demonstrate the theoretical design and experimental implementation of a polarization-multiplexed metasurface-based all-optical neural network to perform various complex object recognition tasks, such as recognizing handwritten digits and fashion items. The physical network is integrated with CMOS imaging sensors for miniaturized and portable sensing and computing all-in-one chip, which can be easily mass-produced because both CMOS chips and metasurface can be manufactured based on semiconductor processes. Another huge advantage of MDNN is the ability to fully exploit parallel operations of light by using the multiplexing of the metasurface. Many multiplexing schemes of the metasurface, including more polarization channel multiplexing[38], wavelength multiplexing[44], spatial multiplexing[45], and vortex multiplexing[46], can be endowed to the all-optical neural network to expand neural network channels. Moreover, the proposed MDNN has a subwavelength pixel size of 400 nm in the visible range, empowering the effective areal density of neurons of $6.25\times10^{6}/mm^{2}$ for a single channel which will be further boosted by the combination with multiplexing. Though our fabricated on-chip MDNN has only one hidden layer, the simplest neural network, it is sufficient to demonstrate the classification of dual targets within two channels (Figures 5 and S4). To obtain higher recognition accuracy and more complex recognition characteristics, multi-layer meta-neurons can be precisely fabricated by overlay lithography[47,48]. To verify the feasibility, we have designed and simulated the multilayer cladding metasurfaces (see Supplementary Note 2 for more details) as well as an MDNN framework with five hidden layers with 280×280×5 meta-nuerons (Figure S7). As a new class of deep learning chips for parallel processing, the pre-trained metasurface devices combined with optical imaging sensors enable to perform complex functions as simply as the human eye, and may open up a new generation of optical multi-skilled AI chips.

## Methods

### Training of the MDNN.

Our MDNN architectures were implemented using Python (v3.6.12) and TensorFlow (v2.1.0, Google Inc.) on a server (GeForce RTX 2080 Ti graphical processing unit (GPU, Nvidia Inc.) and Intel(R) Core (TM) i9-10980XE CPU @3.00 GHz central processing unit (CPU, Intel Inc.) with 128 GB of RAM, running the Windows 10 operating system (Microsoft)). We trained each network in the multi-channel MDNN individually, using the cross-entropy loss as a loss function, which is often used in machine learning for object classification, to maximize the signal in the target region. The neurons in each layer of the network, i.e., phases of the meta-units, are updated by a stochastic gradient descent algorithm. We used the MNIST and Fashion-MNIST datasets for training with a training batch size of 10 and a learning rate of 0.1. The number of neurons per layer in scalar simulations and vector simulations for training was set to be $28 \times 28$, while the number of neurons in a single layer in the experiments for training was set to be $280 \times 280$. The ideal mapping function between the input and output planes was achieved after 2000, 600, and 500 epochs, respectively, and each network took about a few minutes to tens of minutes to train. Further, we trained five hidden layers with 280×280 neurons per layer (see Supplementary Note 8). After training, the correctness of the network is verified by the Rayleigh-Sommerfeld diffraction calculation program using MATLAB.

**Sample Fabrication.**

The MDNN sample was fabricated mainly by two processes, namely the fabrication of the metasurface and the integration with a CMOS imaging sensor, the first of which in turn consists of deposition, overlay EBL, lift-off and atomic layer deposition (ALD), among others. First, after EBL (Raith-$150^{\text{two}}$) patterning of a layer of poly methyl methacrylate (PMMA) resist (950k-8%), aurum (Au) deposition and lift-off, overlay markers were defined on a quartz substrate. Subsequently, a PMMA resist layer was again coated, and after precise overlay exposure using Au markers, development, deposition and lift-off, binary aluminum structure of the input signal to be identified was obtained. A 100 nm spacer protecting the Aluminum layer was obtained by exposing hydrogen silsesquioxane (HSQ). Next, the sample was coated with a 600 nm PMMA again, and the overlay marks were used to define the multiplexed meta-units pattern. After development (1 min in 1:3 MIBK:IPA solution and 1 min in IPA at -18 ℃), an ALD system with $TiCl_4$ precursor was used to deposit amorphous $TiO_2$ onto the resist. Then, the $TiO_2$ film on the top of the sample was etched by ion beam etching (IBE) and the PMMA resist was stripped by reactive ion etching (RIE). Finally, we manufacture metasurface-based diffractive neural networks sample on a Sony IMX686 CMOS chip with an imaging screen of $8.64{\times}6.46\ mm^2$and a pixel of $0.8\ um$. The most critical step in this process is to ensure that the distance between the metasurface sample and the imaging surface is sufficiently precise. Note that the diffraction distance between the hidden layers is 100 μm. Therefore, we cut an optically clear adhesive (OCA) with a thickness of 100 μm into the desired shape so that the metasurface is tightly bonded to the image sensor.

**Experiment setup.**

The experimental setup for the MDNN characterization is presented in Figure S9. A laser diode emitting at 532 nm (Thorlabs CPS532) was utilized as the input light. A linear polarizer (LP) was used to create the desired polarizations. The light is then directed onto the metasurface and imaged on a charge coupled device (CCD) camera through a 100× objective lens. Videos of the MDNN focusing effect with diffraction distance are obtained by the movement of a stepper motor. Since the metasurface is integrated onto the CMOS imaging sensor, the output images of the experiment in Figure 5 are collected directly by the image sensor on the CMOS chip (Sony IMX686).

**Acknowledgments**

The authors thank Dr. Q. Song for discussion of the diffraction algorithm, Mr. Z. Xu for participation in the construction of the MDNN framework, and Mr. P. Wang for providing CMOS chips. The authors acknowledge the financial support by the National Natural Science Foundation of China (Grant No. 52005175, 51722503), Natural Science Foundation of Hunan Province of China (Grant No. 2020JJ5059) and Shenzhen Science and Technology Program (Grant No. RCBS20200714114855118).

**Author contributions:**

Y. Hu, H. Duan, and X. Luo proposed the idea. X. Luo, Y. Hu conceived and carried out the design and simulation. X. Luo, X. Li, J. Lai prepared the metasurface samples. X. Luo, X. Ou and Y. Hu performed the CMOS integration and testing of the MDNN. X. Luo, Y. Hu, N. Liu, and H. Duan analyzed the results and co-wrote the paper. All the authors discussed the results and commented on the manuscript. H. Duan initiated and supervised the project.

**Competing interests:** The authors declare that they have no competing interests.